\documentclass[11pt,letter,aps,preprintnumbers,amsmath,amssymb,superscriptaddress,nofootinbib]{revtex4}
  
\usepackage{graphicx,subfigure,multirow}
 \usepackage{longtable}
\usepackage{dcolumn}
\usepackage{bm}
\usepackage[all]{xy}
\usepackage{color}
 \usepackage{graphicx}
\usepackage[usenames,dvipsnames]{xcolor}
\usepackage[unicode=true,pdfusetitle,
 bookmarks=true,bookmarksnumbered=false,bookmarksopen=false,citecolor=Turquoise,
 breaklinks=false,pdfborder={0 0 1},backref=false,colorlinks=true,pdfpagemode=FullScreen]
 {hyperref}
 
\newcommand{\iab}{\text{ab}^{-1}}

\preprint{Contribution to Snowmass 2021: EF01, TF07}

\newcommand\snowmass{
\begin{center}
  \rule[-0.3in]{\hsize}{0.01in}\\
  \rule{\hsize}{0.01in}\\
  Submitted to the Proceedings of the US Community Study\\
  \vspace{-0.1in}
  on the Future of Particle Physics (Snowmass 2021)\\
  \vspace{-0.1in}
  \rule{\hsize}{0.01in}\\
  \rule[+0.3in]{\hsize}{0.01in}
\end{center}
\vspace{-1cm}
}

\begin{document}

\vspace*{-1.5cm}
\snowmass

\title{Directly Probing the CP-structure of the Higgs-Top Yukawa at HL-LHC and Future Colliders} 
 
\medskip
 
\author{Rahool Kumar Barman}
\affiliation{Department of Physics, Oklahoma State University, Stillwater, OK, 74078, USA}

\author{Morgan E. Cassidy}
\affiliation{Department of Physics and Astronomy, University of Kansas, Lawrence, Kansas 66045~ U.S.A.}

\author{Zhongtian Dong}
\affiliation{Department of Physics and Astronomy, University of Kansas, Lawrence, Kansas 66045~ U.S.A.}

\author{Dorival Gon\c{c}alves}
\affiliation{Department of Physics, Oklahoma State University, Stillwater, OK, 74078, USA}

\author{Jeong Han Kim}
\affiliation{Department of Physics, Chungbuk National University, Cheongju, 28644, Korea}

\author{Felix Kling}
\affiliation{Deutsches Elektronen-Synchrotron DESY, Notkestr. 85, 22607 Hamburg, Germany}

\author{Kyoungchul Kong}
\affiliation{Department of Physics and Astronomy, University of Kansas, Lawrence, Kansas 66045~ U.S.A.}

\author{Ian M. Lewis}
\affiliation{Department of Physics and Astronomy, University of Kansas, Lawrence, Kansas 66045~ U.S.A.}

\author{Yongcheng Wu}
\affiliation{Department of Physics, Oklahoma State University, Stillwater, OK, 74078, USA}

\author{Yanzhe Zhang}
\affiliation{Department of Physics and Astronomy, University of Kansas, Lawrence, Kansas 66045~ U.S.A.}

\author{Ya-Juan Zheng}
\affiliation{Department of Physics and Astronomy, University of Kansas, Lawrence, Kansas 66045~ U.S.A.}


\begin{abstract}
\vspace{0.2cm}
\begin{center}{\large \textbf{Executive Summary}}\end{center} 
\vspace{-0.3cm}
Constraining the Higgs boson properties is a cornerstone of the LHC program and future colliders. In this Snowmass contribution, we study the potential to directly probe the Higgs-top CP-structure via the $t\bar{t}h$ production at the HL-LHC, 100 TeV FCC and muon colliders. We find the limits on the CP phase ($\alpha$) at 95\% CL are $|\alpha| \lesssim 36^\circ$ with dileptonic $t\bar t (h\to b\bar b) $ and $|\alpha| \lesssim 25^\circ$ with combined $t\bar t (h\to \gamma\gamma) $ at the HL-LHC. The 100 TeV FCC brings a significant improvement in sensitivity with $|\alpha| \lesssim 3^\circ$ for the dileptonic $t\bar t (h\to b\bar b) $, due to the remarkable gain in the signal cross-section and the increased luminosity. At future muon colliders, we find that the bounds with semileptonic $t\bar t (h\to b\bar b) \nu\bar\nu$ are $|\alpha| \lesssim 9^\circ$ for 10 TeV and $|\alpha| \lesssim 3^\circ$ for 30 TeV, respectively.
\end{abstract}

\maketitle

\vspace*{1cm}
    \begin{center}
    \begin{tabular}{c|c|c|c}
    \hline\hline
         ~Bounds on $\alpha$ at 95\% CL ($\kappa_t=1$)~~ & Channel & Collider & ~Luminosity~ \\      
            \hline\hline
         ~~~$|\alpha| \lesssim 36^\circ$~\cite{Goncalves:2021dcu} ~~
         & dileptonic $t\bar t(h \to b\bar b)$
         & HL-LHC  
         & 3 ab$^{-1}$ \\
            \hline
         $|\alpha| \lesssim 25^\circ$ \cite{Barman:2021yfh} 
         & ~~$t\bar t(h \to \gamma\gamma)$ combination ~~
         & HL-LHC  
         & 3 ab$^{-1}$ \\
            \hline
         $|\alpha| \lesssim 3^\circ$ \cite{Goncalves:2021dcu}  
         & dileptonic $t\bar t(h \to b\bar b)$
         & 100 TeV FCC 
         & 30 ab$^{-1}$ \\
            \hline
         $|\alpha| \lesssim 9^\circ$ \cite{ToAppear} 
         & semileptonic $t\bar t(h \to b\bar b)$
         & 10 TeV $\mu^+\mu^-$  
         & 10 ab$^{-1}$ \\
            \hline
         $|\alpha| \lesssim 3^\circ$  \cite{ToAppear} 
         & semileptonic $t\bar t(h \to b\bar b)$
         & ~~30 TeV $\mu^+\mu^-$ ~~
         & 10 ab$^{-1}$ \\
    \hline\hline
    \end{tabular}
    \end{center}




\newpage
\section{Introduction}

Beyond the Standard Model CP effects to the top-quark Yukawa can be parameterized as
\begin{align}
    \mathcal{L} \supset -\frac{m_t}{v}\kappa_t \bar{t}\left(\cos\alpha+i\gamma_5\sin\alpha\right)t \, h\,,
\end{align}
where $\alpha$ is the CP-phase, $\kappa_t$ is a real number that controls the interaction strength, and ${v=246}$~GeV. Following this parametrization, the SM is described with $\kappa_t=1$ and $\alpha=0$. In contrast, a purely CP-odd interaction would display $\alpha=\pi/2$.

Numerous kinematic observables have been proposed in the literature to  probe the CP-structure of the Higgs-top interaction~\cite{Ellis:2013yxa,Boudjema:2015nda,Buckley:2015vsa,Buckley:2015ctj,Demartin:2014fia,Gritsan:2016hjl,Goncalves:2016qhh,Mileo:2016mxg,AmorDosSantos:2017ayi,Azevedo:2017qiz,Li:2017dyz,Goncalves:2018agy,ATLAS:2018mme,CMS:2018uxb,Ren:2019xhp,Bortolato:2020zcg,Cao:2020hhb,Martini:2021uey,Goncalves:2021dcu,Barman:2021yfh,Bahl:2022yrs}. Some illustrative examples are the transverse momentum of the Higgs boson $p_{Th}$, the invariant mass of the top-quark pair $m_{tt}$, and the angle between the beam direction and the top-quark in the $t\bar{t}$ center of mass frame $\theta^*$ also known as Collins-Soper angle. These variables are sensitive to the squared terms $\cos^2\alpha$ and $\sin^2\alpha$, being CP-even. In particular, they are insensitive to the sign of the CP-phase. CP-odd observables can be constructed from  antisymmetric tensor products that demand four four-momenta. Owning to the top-quark short lifetime, the top-quark polarization is passed to its decay products. Thus, it is possible to construct such tensor product with the top pair and their decay products, such as $\epsilon(p_t,p_{\bar t},p_i,p_k)\equiv \epsilon_{\mu\nu\rho\sigma}p_t^\mu p_{\bar t}^\nu p_i^\rho p_k^\sigma$~\cite{Boudjema:2015nda, Mileo:2016mxg, Goncalves:2018agy}. This tensor product can be simplified to $\vec p_t \cdot (\vec p_i\times \vec p_k)$ in the $t\bar t$ CM frame, granting the definition of azimuthal angle differences that are odd under CP transformation
\begin{align}
        \!\!\! \Delta \phi_{ik}^{t\bar{t}} \!=\!
    \text{sgn} \left[\vec{p}_{t} \!\cdot\! (\vec{p}_{i} \!\times\! \vec{p}_{k})\right] 
   \arccos \!\left( \frac{\vec{p}_{t} \!\times\! \vec{p}_{i}}{|\vec{p}_{t} \!\times\! \vec{p}_{i}|} \!\cdot\! \frac{\vec{p}_{t} \!\times\! \vec{p}_{k}}{|\vec{p}_{t} \!\times\! \vec{p}_{k}|}\right)\!.
    \label{eq:CP_odd_obseravable}
\end{align}

\begin{figure*}[!t]
   \centering
   \includegraphics[width=0.48\textwidth]{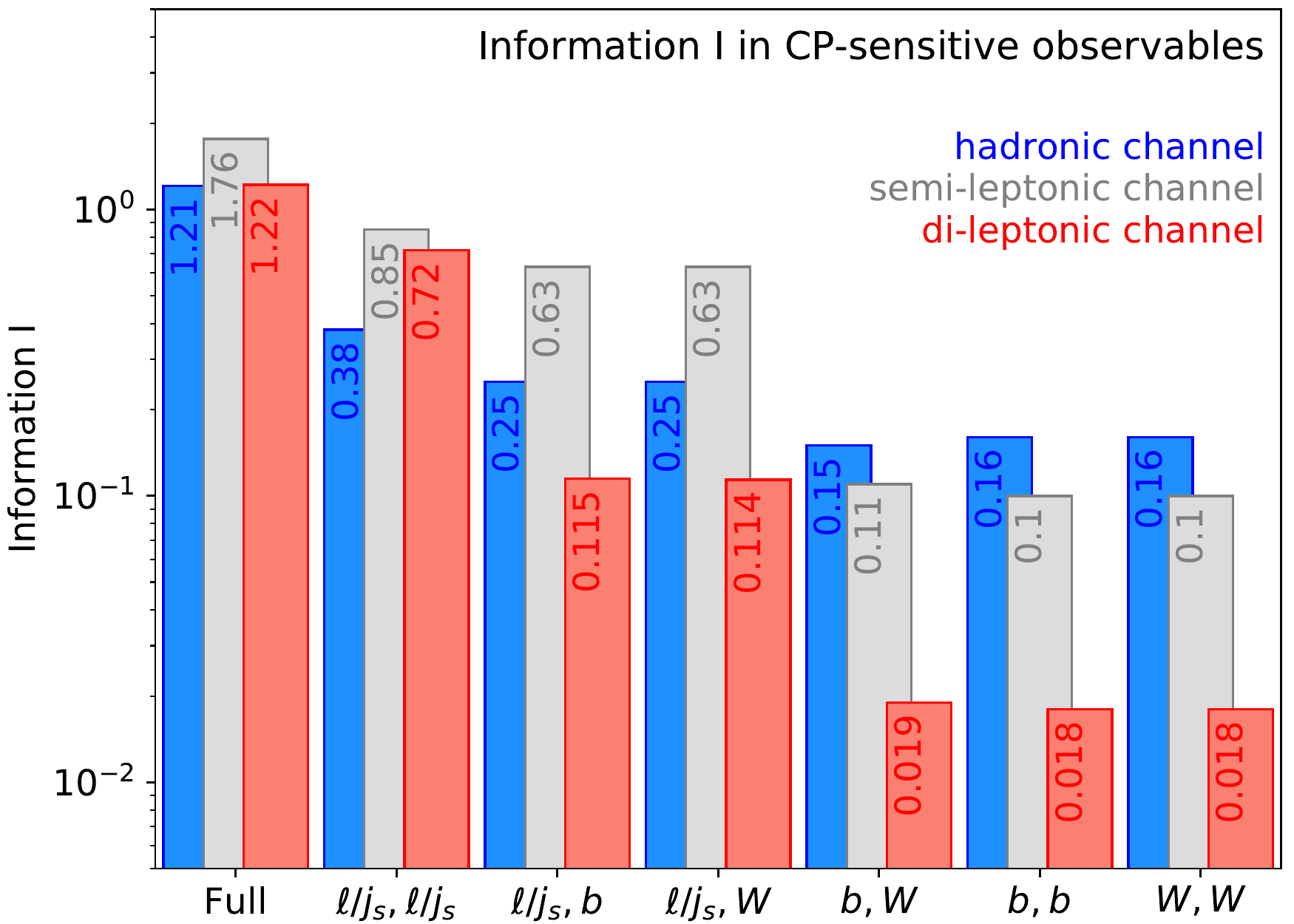}
   \includegraphics[width=0.48\textwidth]{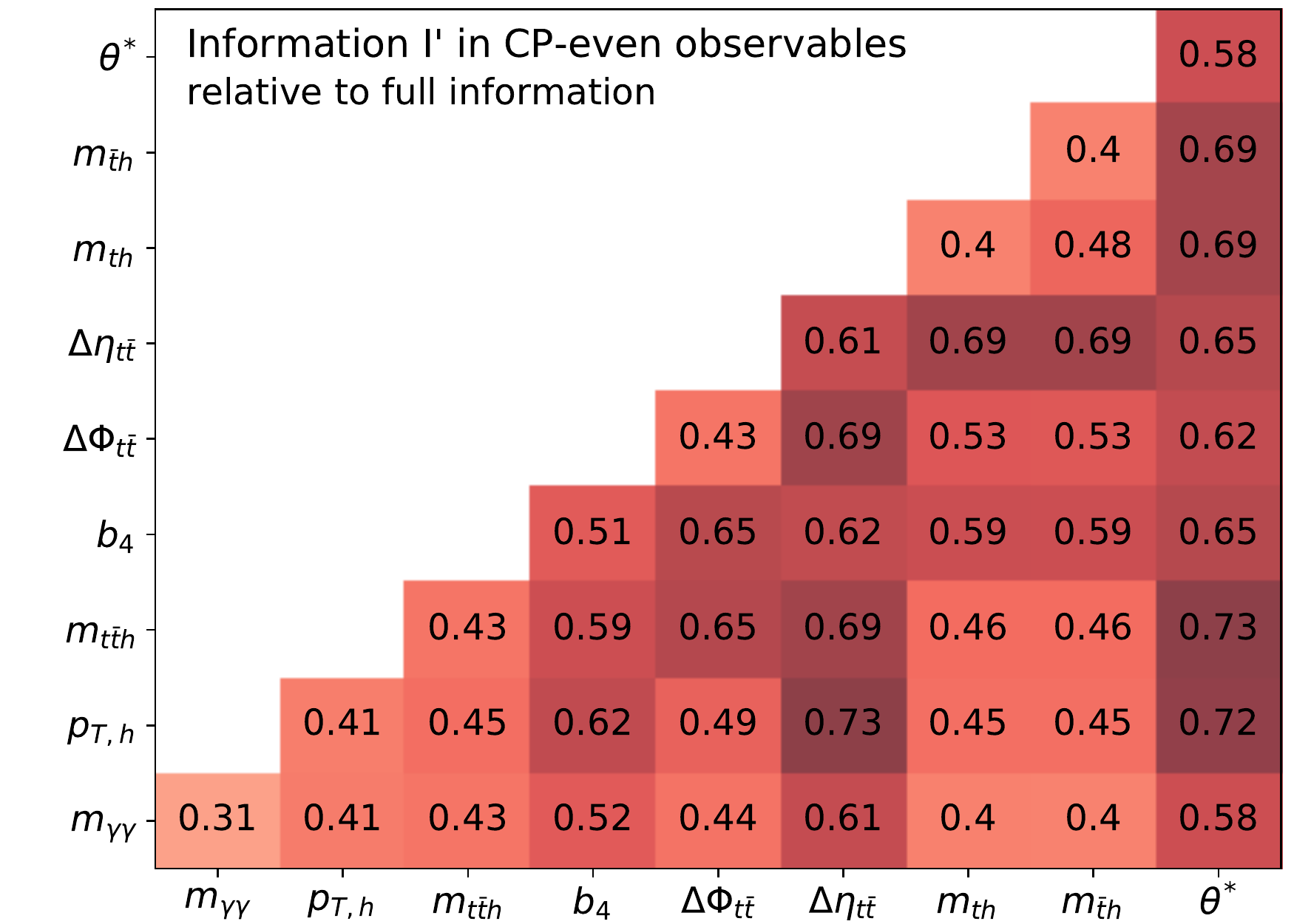}
    \caption{Left panel: Fisher information $I$ from the CP-odd observables for the di-leptonic (red), semi-leptonic (gray), and hadronic (blue) top pair final state. Right panel: Modified Fisher information $I'$ sensitive to the CP-even observables. These results are from Ref.~\cite{Barman:2021yfh}.}
    \label{fig:information}
\end{figure*}

Given the complex multiparticle $t\bar th$ phase space, it is enlightening to quantify and compare how much information on the CP-phase $\alpha$ is available using the distinct CP-even or CP-odd observables. Ref.~\cite{Barman:2021yfh} addressed this task, using the Fisher information as a convenient metric. The results are reproduced in Fig.~\ref{fig:information}. The $\Delta \eta_{t\bar t}$ and $\theta^*$ are the most sensitive CP-even observables, carrying approximately 60\% of the full information. Successively adding further observables augment the information. This highlights that it is crucial to perform a multivariate analysis to maximize the CP sensitivity. Efficient kinematic reconstruction methods will play a crucial role in this task. In particular, it will grant the reconstruction of the top-quark pairs required for the CP-odd observables in Eq.~\ref{eq:CP_odd_obseravable}. This is notably challenging for the dileptonic top-pair final state, that presents two missing neutrinos.

\section{HL-LHC}

\subsection{$pp\to t\bar{t}h(\gamma\gamma)$}

ATLAS and CMS high-luminosity LHC (HL-LHC) projections indicate that the $pp\to t\bar th$ channel, in the diphoton $h\to \gamma\gamma$  final state, will display dominant sensitivities to the top-quark Yukawa strength $\kappa_t$~\cite{Cepeda:2019klc}. Despite the limited statistics, the diphoton final state analysis strongly benefits from controlled backgrounds from the side-bands. Exploring this fact, Ref.~\cite{Barman:2021yfh} shows that a combination of machine learning techniques and efficient kinematic reconstruction methods can boost new physics sensitivity on the CP-phase $\alpha$, effectively probing the complex $t\bar{t}h$ multi-particle phase space. Special attention is committed to top quark polarization observables, uplifting the study from a raw rate to a polarization analysis.

\begin{figure}[t!]
    \centering
    \includegraphics[scale=0.3]{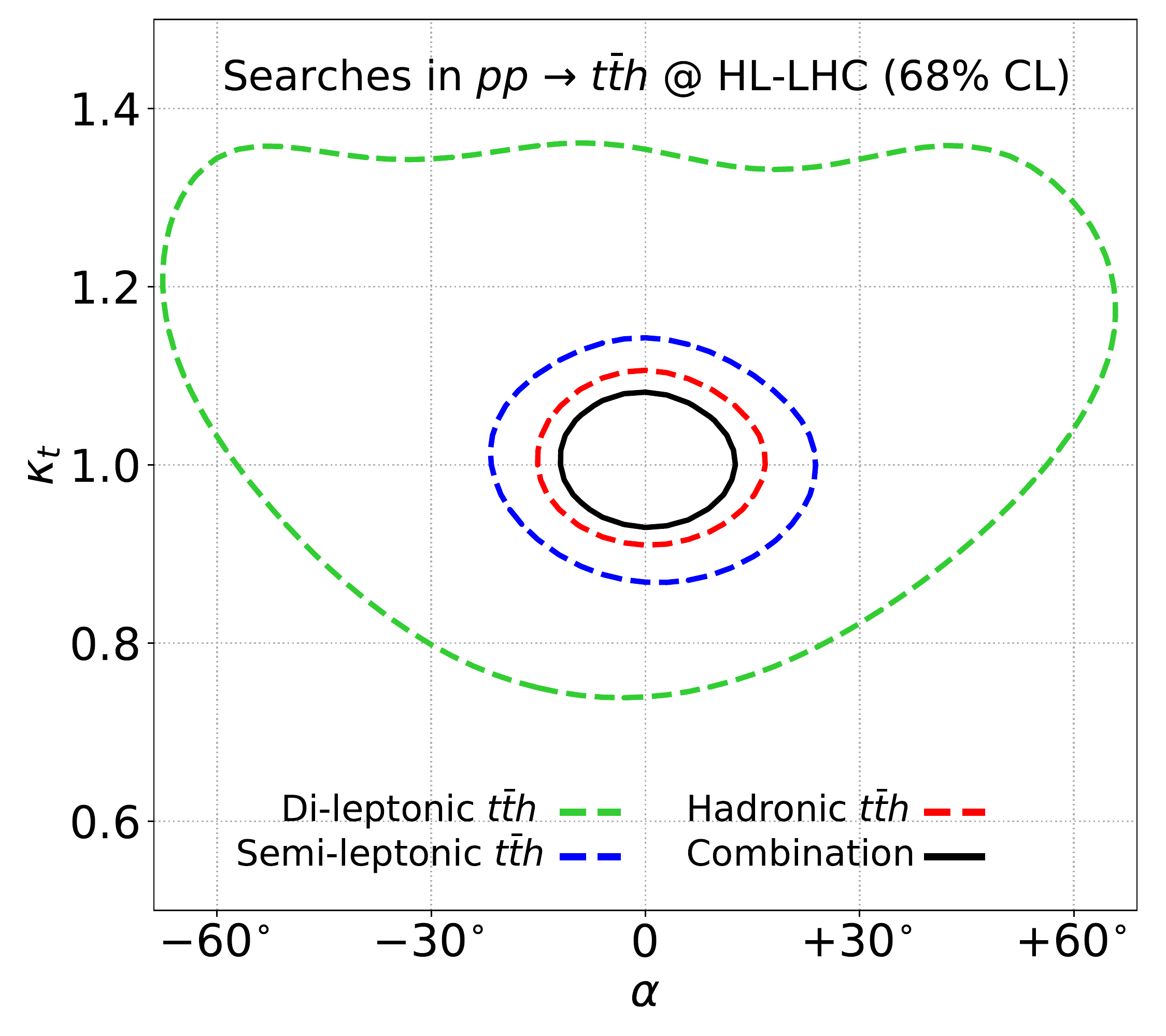}\hspace*{-0.1cm}
    \includegraphics[scale=0.3]{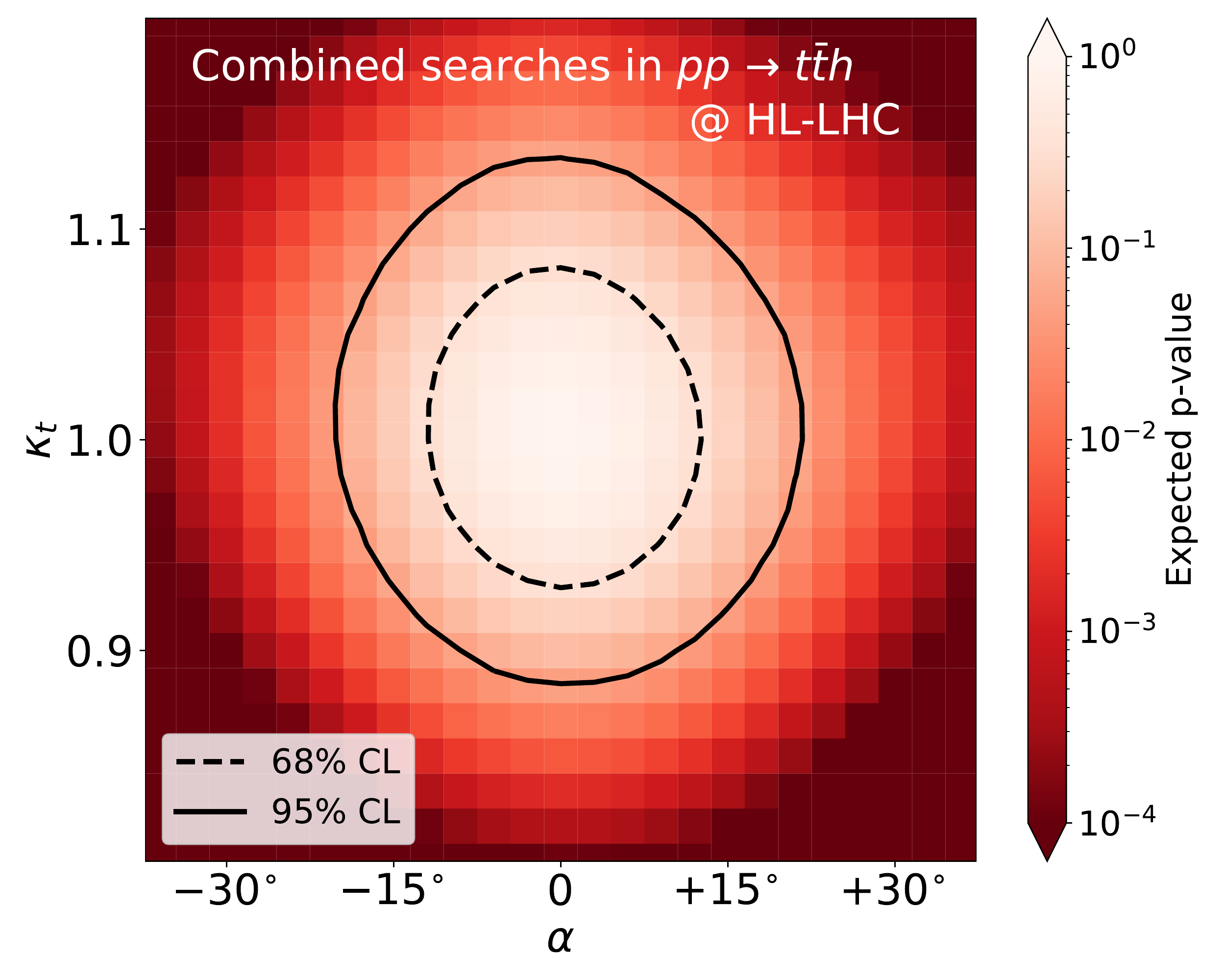} \hspace*{-0.1cm}
    \caption{Projected sensitivity in the $(\alpha,\kappa_t)$ plane. In the left panel we show the projected $68\%$ CL contours from direct Higgs-top searches in the semi-leptonic~(blue), di-leptonic~(green), hadronic~(red) $t\bar{t}h$ channels, and their combination~(black), considering all input observables. The right panel shows the projected $68\%$~CL (dashed) and $95\%$~CL (solid) contours from the combination of the three channels. The color palette illustrates the expected p-value of the estimated log-likelihood ratio.  The projections are derived for 14 TeV LHC assuming $\mathcal{L} = 3~\iab$. The results are from Ref.~\cite{Barman:2021yfh}.
     } \label{fig:combined_excl_limits} 
\end{figure}

In Fig.~\ref{fig:combined_excl_limits}, we summarize the projected sensitivity in the $(\alpha,\kappa_t)$ plane. In the left panel we show the projected $68\%$ CL contours from direct Higgs-top searches in the semi-leptonic~(blue), di-leptonic~(green), hadronic~(red) $t\bar{t}h$ channels, and their combination~(black), considering all input observables. The right panel shows the projected $68\%$~CL (dashed) and $95\%$~CL (solid) contours from the combination of the three channels, considering all input observables. The color palette illustrates the expected p-value of the estimated log-likelihood ratio. The projections are derived for 14 TeV LHC assuming $\mathcal{L} = 3~\iab$.

\subsection{$pp\to t\bar{t}h(b\bar{b})$}

\begin{figure*}[!t]
    \centering
    \includegraphics[width=\textwidth]{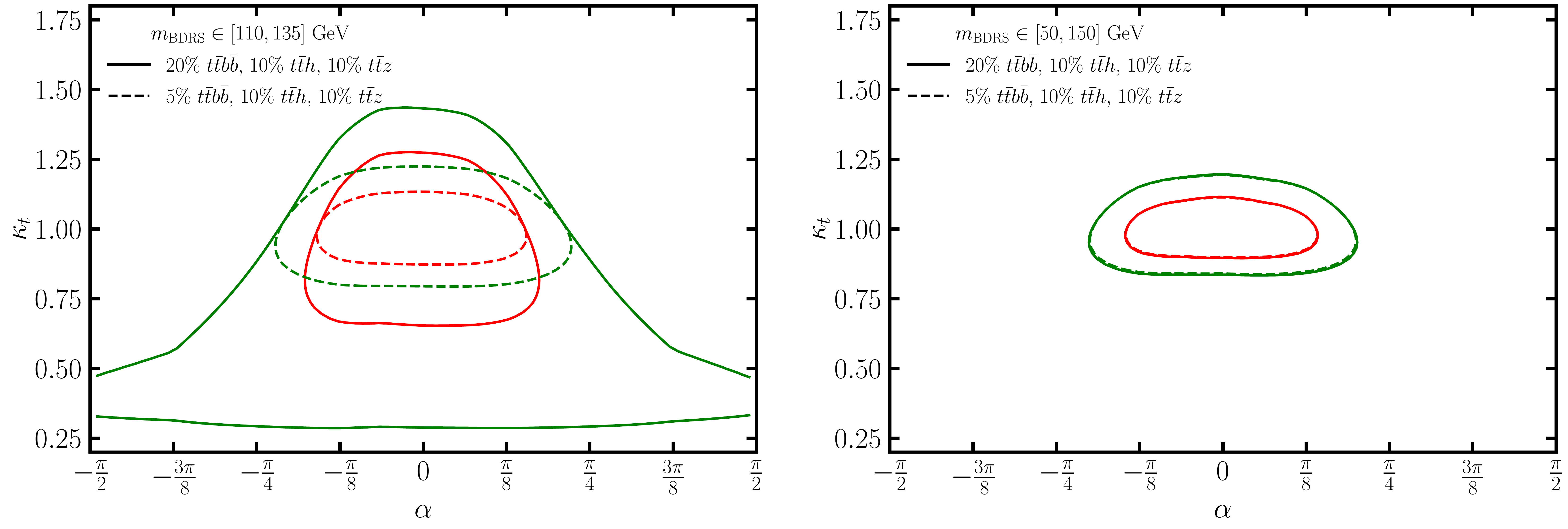}
    \caption{The exclusion at 68\% (red) and 95\% (green) CL in the $\alpha$-$\kappa_t$ plane at the 14~TeV LHC with $3\,{\rm ab}^{-1}$ for a narrow (left) and wide (right) mass window.
    20\% systematics (5\%) for $t \bar t b \bar b$ is assumed in solid (dotted) curves, while 10\% systematics is used for both $t\bar t Z$ and $t \bar th$. The results are from Ref.~\cite{Goncalves:2021dcu}.
    }
    \label{fig:alpha_kt_14TeV}
\end{figure*}

We perform a similar analysis with the Higgs decay to $b\bar b$ and two top quarks to dilepton (see Ref.~\cite{Goncalves:2021dcu} for details), which provides the extra background suppression as well as a better probe to the top polarization, using the charged leptons. The larger spin analyzing power associated with the charged leptons results in the stronger CP-violation observables, such as $\Delta\phi_{\ell\ell}^{t\bar{t}}$, strengthening the CP-sensitivity.
We adopt a binned log-likelihood analysis exploring the Higgs candidate invariant mass profile in the boosted regime, for the signal range ${m_J^\text{BDRS}\in [110,135]\,\rm GeV}$, together with the CP-sensitive observable $\theta^*$.  Since the considered $t\bar{t}h$ channel with $h\to b\bar{b}$ typically encounters a large $t\bar{t}b\bar{b}$ background, which has a significant uncertainty~\cite{ATLAS:2017fak,CMS:2018hnq}, the  final result displays relevant correlation with the considered background uncertainties. To estimate this effect, we derive the new physics sensitivity on the $(\alpha,\kappa_t)$ plane for two scenarios. In the first case, we assume that  $t\bar{t}b\bar{b}$ background rate has 20\% of uncertainty, which is included as a nuisance parameter. The magnitude of the considered error is comparable to that used in the current experimental analyses~\cite{ATLAS:2017fak,CMS:2018hnq}. For the second case, we assume an optimistic scenario with 5\% error. The uncertainties on the $t\bar{t}h$ and $t\bar{t}Z$ samples are assumed to be 10\% for both scenarios.
The result of this analysis is presented in the left panel of Fig.~\ref{fig:alpha_kt_14TeV}. We obtain that the CP-mixing angle can be constrained to $|\alpha|\lesssim 32^\circ$ at 68\%~CL at the HL-LHC for both scenarios. At the same time, we find that the sensitivity from $\kappa_t$ to the systematic error is more pronounced. While in the first scenario we can constrain the top Yukawa strength to $\delta\kappa_t \lesssim 0.3$, the more optimistic case leads to  $\delta\kappa_t\lesssim 0.15$.

To illustrate how to reduce the systematics for $t\bar t b\bar b$ in a realistic measurement, we enlarge the mass range of the Higgs candidate to ${m_J^{\rm BDRS}\in [50,150]\,\rm GeV}$.
In this case, the events outside the Higgs peak, which mainly arise due to $t\bar t b\bar b$ production, can be used together with the shape of $m_J^{\rm BDRS}$ distribution of $t\bar tb\bar b$ from MC simulation within the binned log-likelihood method. Fitting to a broader range of $m_J^{\rm BDRS}$, we obtain a better control of the uncertainties of $t\bar tb\bar b$ and show the results in the right panel of Fig.~\ref{fig:alpha_kt_14TeV}. We find that this analysis depletes the influence of the systematic uncertainties, leading to similar results for the two considered systematic uncertainty scenarios. The obtained limits are $|\alpha|\lesssim26^\circ~(36^\circ)$ and $\delta\kappa_t\lesssim 0.12~(0.2)$ at 68\%~(95\%) CL. Using the wider mass window, the log-likelihood analysis takes full advantage of the shape information of $t\bar t h$ and $t \bar t b \bar b$ events.

\section{100 TeV FCC}

\begin{figure}[t!]
    \centering
    \includegraphics[width=0.5\textwidth]{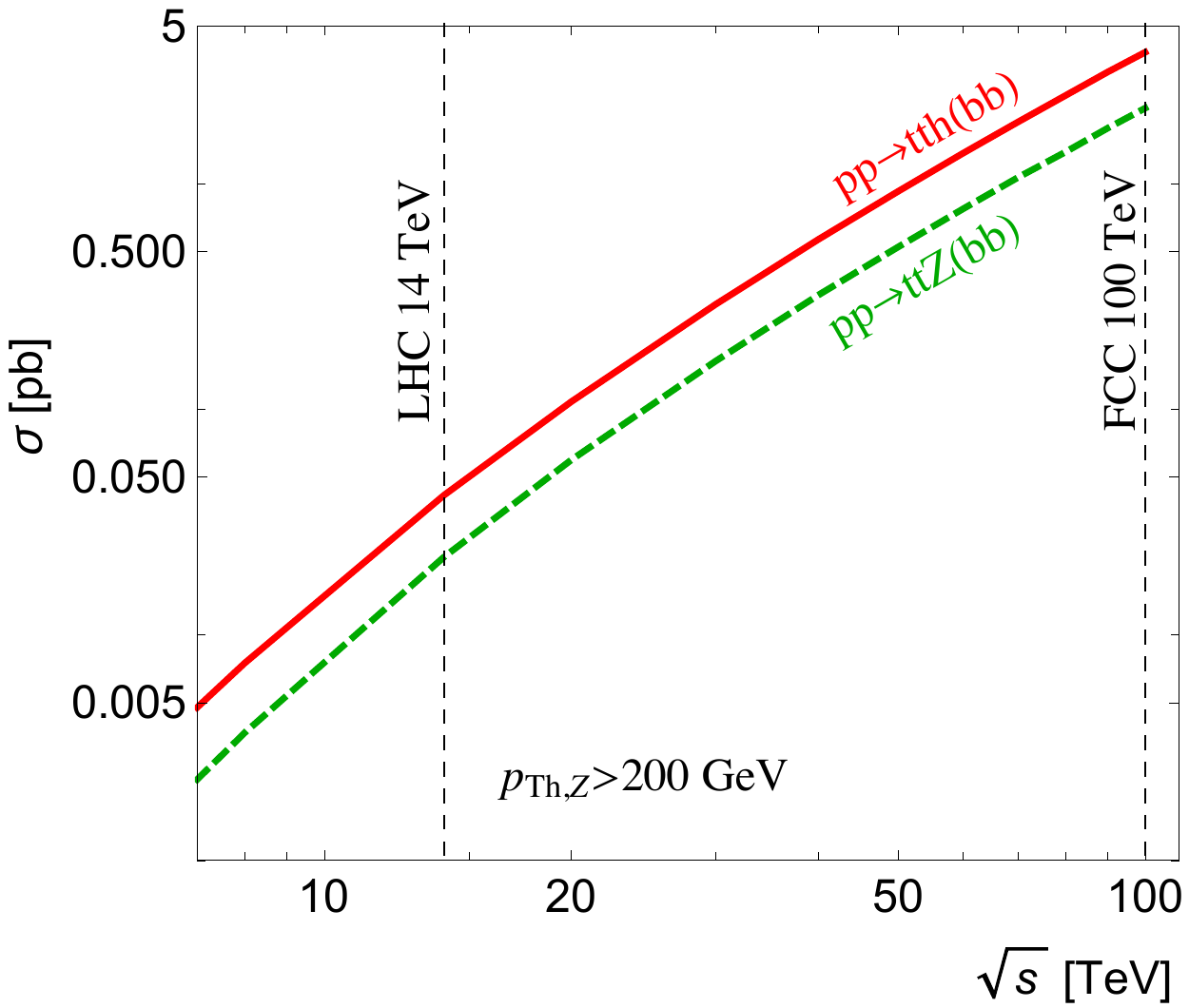}
    \caption{Production cross-section for $pp\to t\bar{t}h$ and $pp\to t\bar{t}Z$ at the parton level as a function of the hadron collider energy. We consider the Higgs and Z bosons in the boosted regime, $p_{Th,Z}>200$~GeV. Their  branching ratios to a bottom-quark pair $\mathcal{BR}(h,Z\to b\bar{b})$ are accounted for. Top quarks are set stable. The results are from Ref.~\cite{Goncalves:2021dcu}.
    \label{fig:ttX-xsection} }
\end{figure}

The Higgs-top CP-phase measurement would obtain remarkable gains at a future 100~TeV collider due to the immensely increased statistics. 
As shown in Fig.~\ref{fig:ttX-xsection}, while the $t\bar{t}(h\to b\bar{b})$ and $t\bar{t}(Z\to b\bar{b})$ processes are phase space suppressed at the 14~TeV LHC, leading to 0.04~pb and 0.02~pb for their cross-sections, the 100~TeV collider would result in one hundred-fold enhancement, with a cross-section of 3.8~pb and 2.1~pb, respectively. Considering the leptonic top pair decay, this corresponds to a significant increase in the number of events (after top quark decays) for the $t\bar{t}h$ signal from $5.8\times 10^3$  at the HL-LHC with 3~ab$^{-1}$ to $5.5\times 10^6$ at 100~TeV with 30~ab$^{-1}$.

\begin{figure*}[t]
    \centering
    \includegraphics[width=\textwidth]{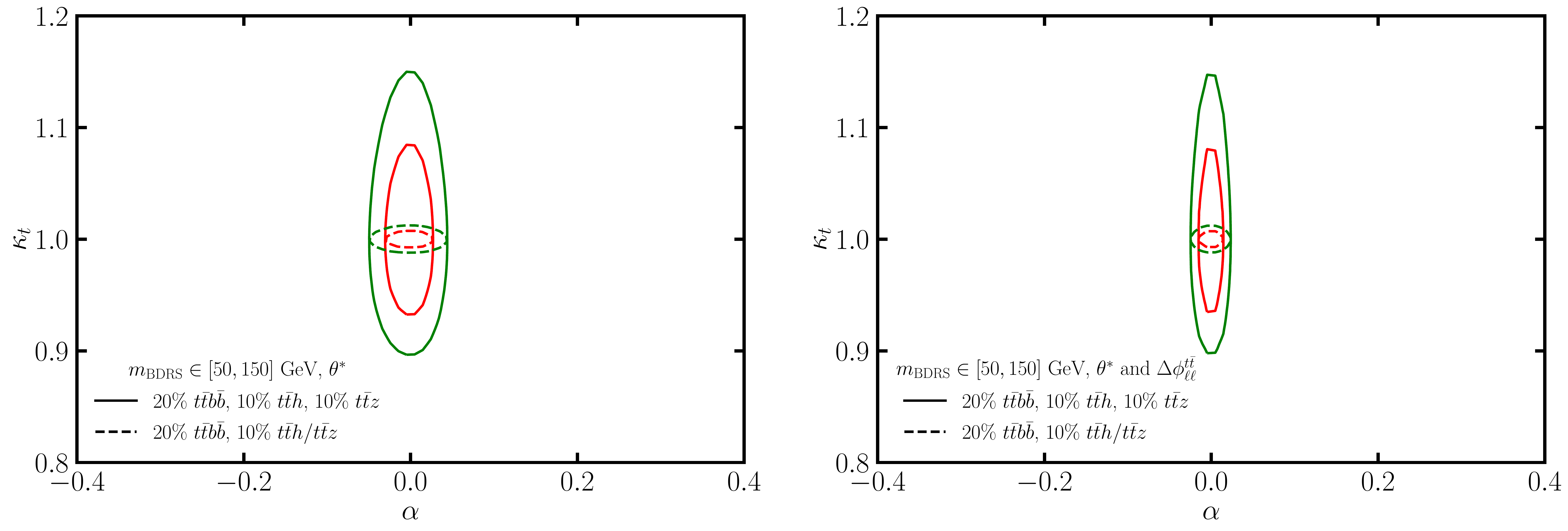}
    \caption{The exclusion at 68\% (red) and 95\% (green) CL limits on the $\alpha$-$\kappa_t$ plane at the 100~TeV FCC with $30\ {\rm ab}^{-1}$ without (left) and with (right)  $\Delta\phi_{\ell\ell}^{t\bar t}$, which is the azimuthal angle between two leptons in the $t\bar t$ rest frame.
    For the solid curves, 10\% systematics is used for both $t\bar th$ and $t\bar tZ$ individually, while for the dashed curves, the uncertainties for $t\bar t h$ and $t\bar t Z$ are assumed to be correlated. 20\% systematics is used for $t\bar tb\bar b$ for both scenarios. The results are from Ref.~\cite{Goncalves:2021dcu}.}
    \label{fig:alpha_kt_100TeV}
\end{figure*}

Performing a similar analysis with the uplifted cross-section and enlarged luminosity, the 100~TeV FCC can boost the sensitivities on ($\alpha$, $\kappa_t$), using the binned log-likelihood method, as summarized in Fig.~\ref{fig:alpha_kt_100TeV}.
We choose a wide mass window, $m_{\rm BDRS} \in [50, 150]$ GeV for better control of the continuum $t\bar{t}b\bar{b}$ background, along with $\theta^\ast$ in the left panel. In both panels, the solid curves correspond to the case with 20\% systematics for $t\bar tb\bar b$ and 10\% systematics for $t\bar t h$ and $t\bar tZ$, while we assume $t\bar t h$ and $t\bar t Z$ uncertainties are correlated for the dashed curves. It is clear that, at high luminosities, the solid curves are limited by the systematic uncertainties. 
However, by assuming that the systematics of $t\bar th$ is correlated with $t\bar tZ$, the precision can be improved, as shown by the dashed curves, which can achieve $\delta\kappa_t\lesssim  1\%$ and $| \alpha| \lesssim 3^\circ$ at 95\%~CL. Finally, extending the analysis to the $(m_{\rm BDRS},\theta^*,\Delta\phi_{\ell\ell}^{t\bar t})$ plane, we find that the CP-odd observable $\Delta\phi_{\ell\ell}^{t \bar t}$ ($\Delta\phi$ between two leptons in the $t\bar t$ rest frame) brings additional improvement on the measurement of $\alpha$ by a factor of 2, $| \alpha| \lesssim 1.5^\circ$, as shown in the right panel of Fig.~\ref{fig:alpha_kt_100TeV}, which highlights the importance of the CP-odd observable in the $t\bar t$ rest frame.

A recent study on ML-inspired reconstruction algorithm can improve our results. As discussed in Refs.~\cite{Goncalves:2018agy,Goncalves:2021dcu}, top quark reconstruction in the dilepton channel plays an important role. These studies show that the endpoints method gives $\sim$74\% efficiency and $\sim$87\% purity at detector-level. 
On the other hand, Ref.~\cite{Alhazmi:2022qbf} shows that the same purity ($\sim$87\%) can be obtained at a higher efficiency using deep neural networks (with 92\% efficiency) and Lorentz Boost Networks (with nearly 100\% efficiency), which would lead to a gain of 24\% and 35\% more events, respectively. 
The ML methods should improve further on the precision measurement of $\kappa_t$ and $\alpha$. More challenging use of ML methods would be direct reconstruction of the kinematic variables that are sensitive to the CP angle \cite{Kim:2021pcz}.

\section{Muon Collider}

\begin{figure}[!t]
    \centering
    \includegraphics[width=0.5\textwidth]{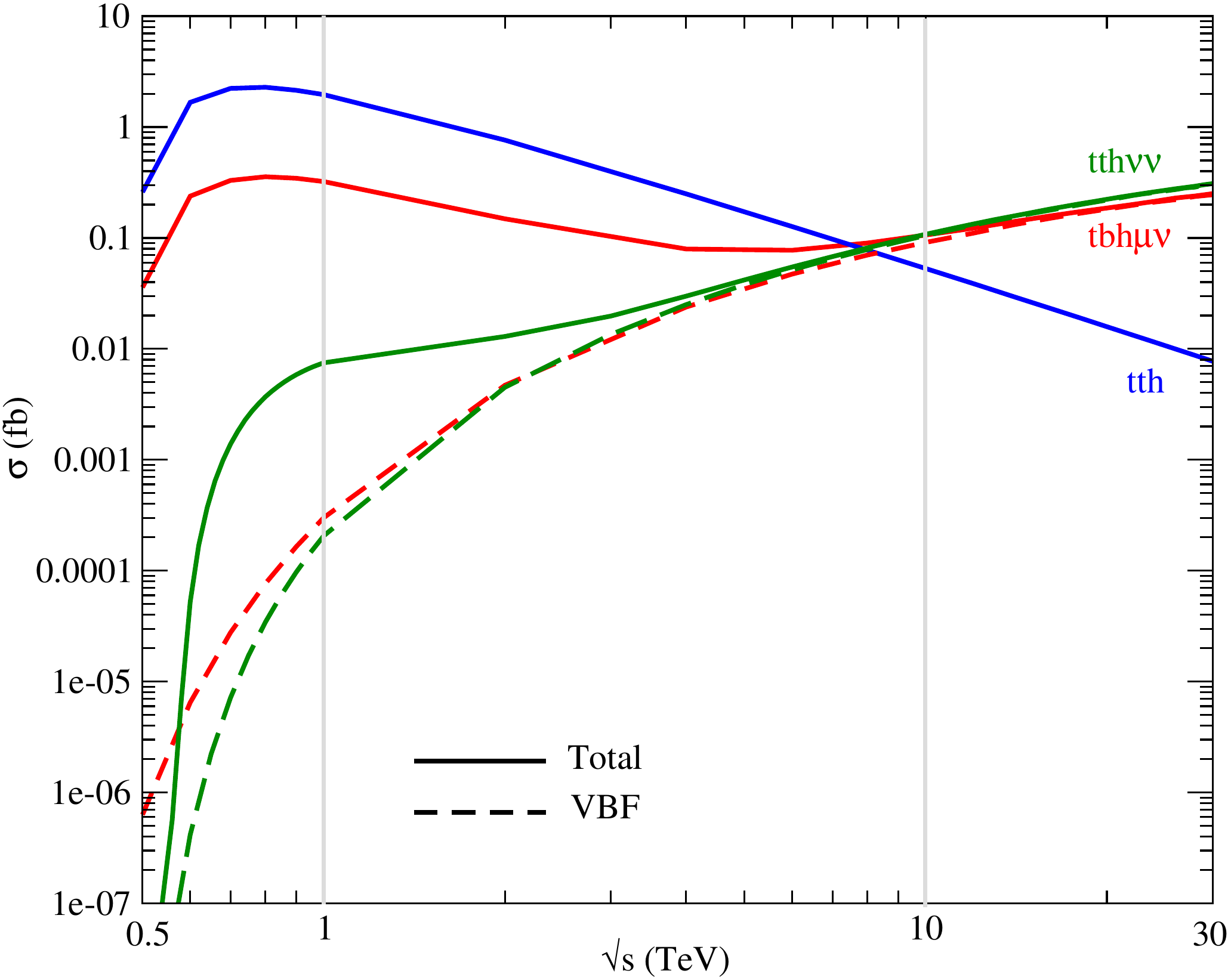}
    \caption{The SM ($\alpha=0$) cross-sections for (blue) $\mu^+\mu^- \to t\bar t h$, (red) $t b h \mu \nu$ and (green) $t\bar t h \nu \bar \nu$ processes as a function of $\sqrt{s}$. VBF contributions are shown as dashed lines and generated by replacing the $\mu^+$ with a positron.  For $tbh\mu\nu$ a cut of $p_T^\mu>10$~TeV is applied.}
    \label{fig:cs}
\end{figure}

 In this section we summarize the results of Ref.~\cite{ToAppear}. Throughout this section we will assume $\kappa_t=1$. We show in Fig.\,\ref{fig:cs} the SM ($\alpha=0$) cross-section versus $\sqrt{s}$ for
  \begin{eqnarray}
  \mu^+\mu^-&\to& t\bar{t}h,\,t\bar{t}h\nu{\bar\nu},\,tbh\mu\nu,~{\rm where}
  \nonumber
  \\
   t\bar{t}h\nu{\bar\nu}&\equiv& t\bar{t}h\nu_\ell{\bar\nu}_\ell\quad(\ell=e~{\rm and} ~\mu)
   \nonumber
   \\
    tbh\mu\nu&\equiv& t\bar{b}h\mu^-\nu_\mu+\bar{t}{b}h\mu^+\bar{\nu}_\mu
  \end{eqnarray}
  with $\sqrt{s}$ from 500 GeV to 30 TeV.  To consistently keep gauge invariance, the vector boson fusion (VBF) subprocess contributions are shown separately as dashed lines for $tbh\mu\nu$ and $t\bar{t}h\nu\bar{\nu}$ by replacing the $\mu^+$ with $e^+$ as adopted in Ref. \cite{Costantini:2020stv}.  All events are generated by implementing the CP-violating model in \texttt{MadGraph5\_aMC@NLO}~\cite{Alwall:2014hca} via \texttt{FeynRules}~~\cite{Christensen:2008py,Alloul:2013bka}.
   
 At low energies, the $t\bar{t}h$ production is dominant, while around $\sqrt{s}\sim7-8$ TeV the $t\bar{t}h\nu\bar{\nu}$ and $tbh\mu\nu$ take over. This can be understood by noting that the $t\bar{t}h$ production only occurs through $s$-channel off-shell $\gamma/Z$ contributions.  Hence, the $t\bar{t}h$ cross-section decreases with $\sqrt{s}$ above threshold, from about 2.0 fb at 1 TeV to $7.8\times 10^{-3}$~fb at 30 TeV.  However, for the $t\bar{t}h\nu\bar{\nu}$ and $tbh\mu\nu$ production, the VBF contribution is dominant for $\sqrt{s}\gtrsim 3$~TeV causing the cross-sections to grow with energy.  For $tbh\mu\nu$ production, there is a VBF diagram where one muon radiates a photon and other a $W$.  For example, a representative process is
 \begin{eqnarray}
 \mu^-\mu^+ \rightarrow \mu^- \bar{\nu} (\gamma^* W^{+*}\rightarrow t\bar{b}h),
 \end{eqnarray}
 where the parenthesis indicates the radiated photon and $W$ fuse into $tbh$.  The massless photon mediator in this case causes a singularity when the final and initial state muons are collinear.  We impose a cut on the transverse momentum of the outgoing muon $p_T^{\mu}>10$ GeV to avoid the singularity and generate numerically stable results\footnote{A proper treatment involves the effective vector boson approximation~\cite{vonWeizsacker:1934nji,Williams:1935dka,Dawson:1984gx,Ruiz:2021tdt}.  For $\sqrt{s}\gtrsim 30$~TeV, all VBF processes should incorporate the effective vector boson approximation to resum the large logs~\cite{Costantini:2020stv}. }.

\begin{figure}[!t]
    \centering
    \includegraphics[width=0.32\textwidth,clip]{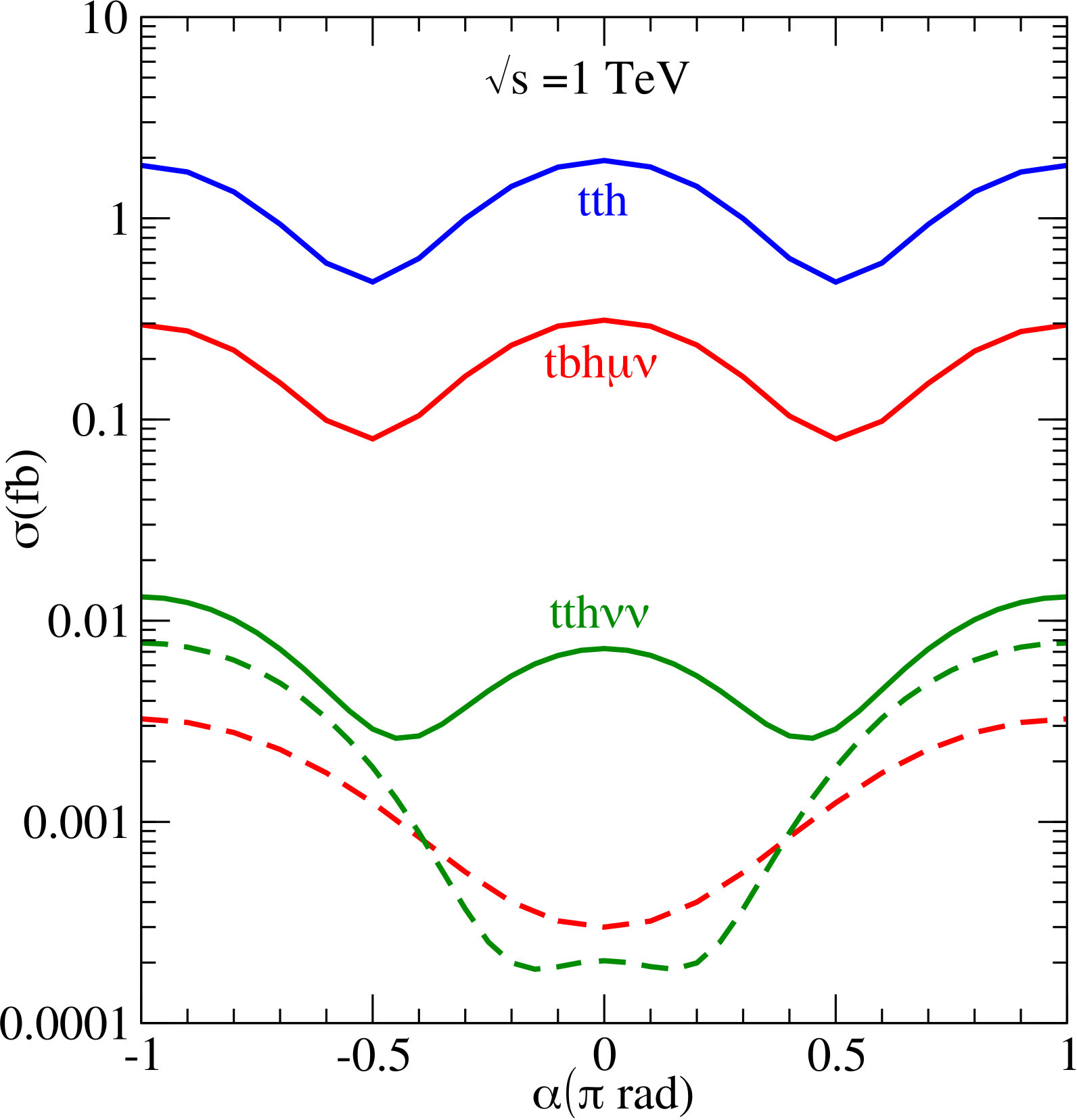}
    \includegraphics[width=0.305\textwidth,clip]{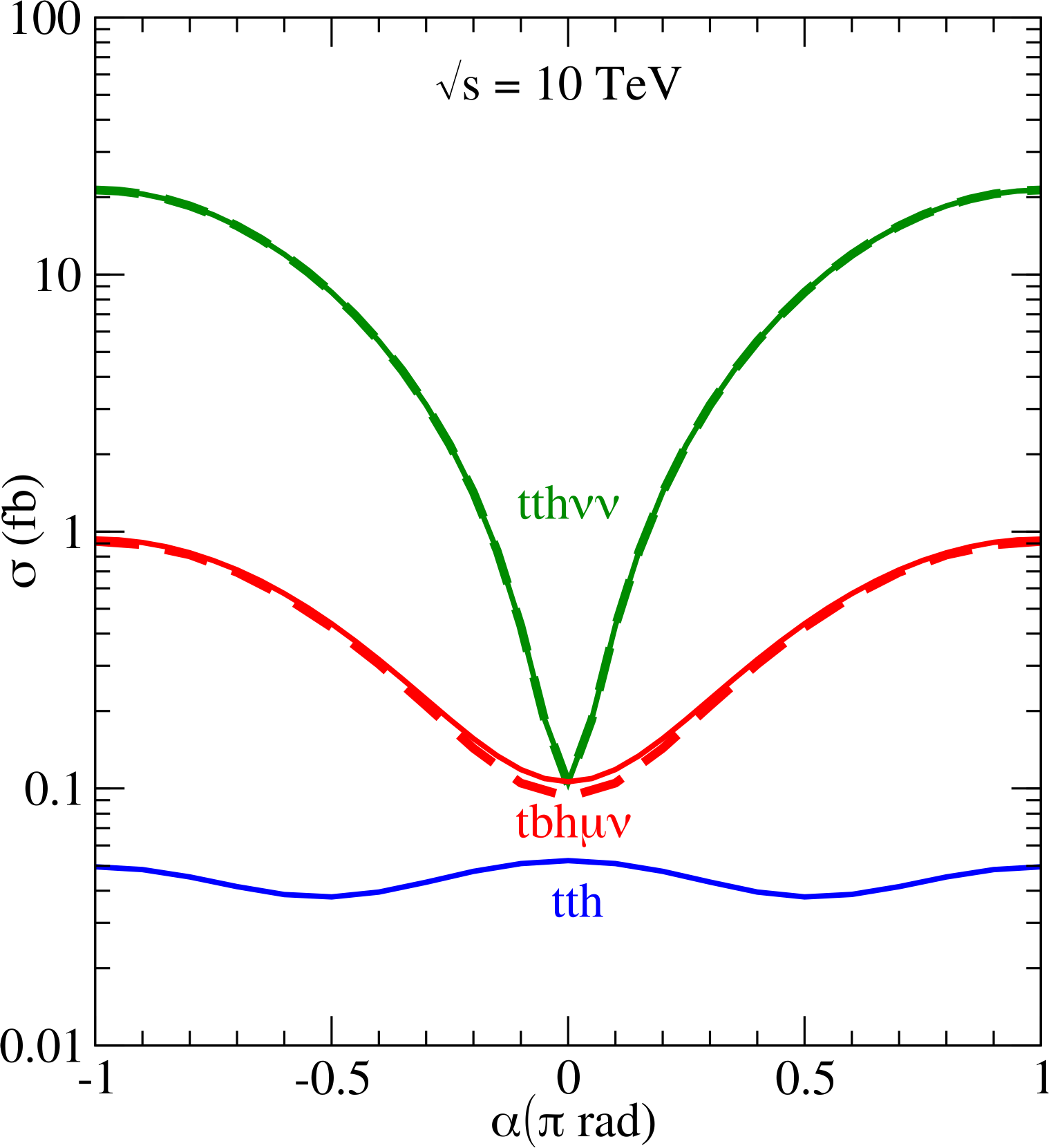}
    \includegraphics[width=0.31\textwidth,clip]{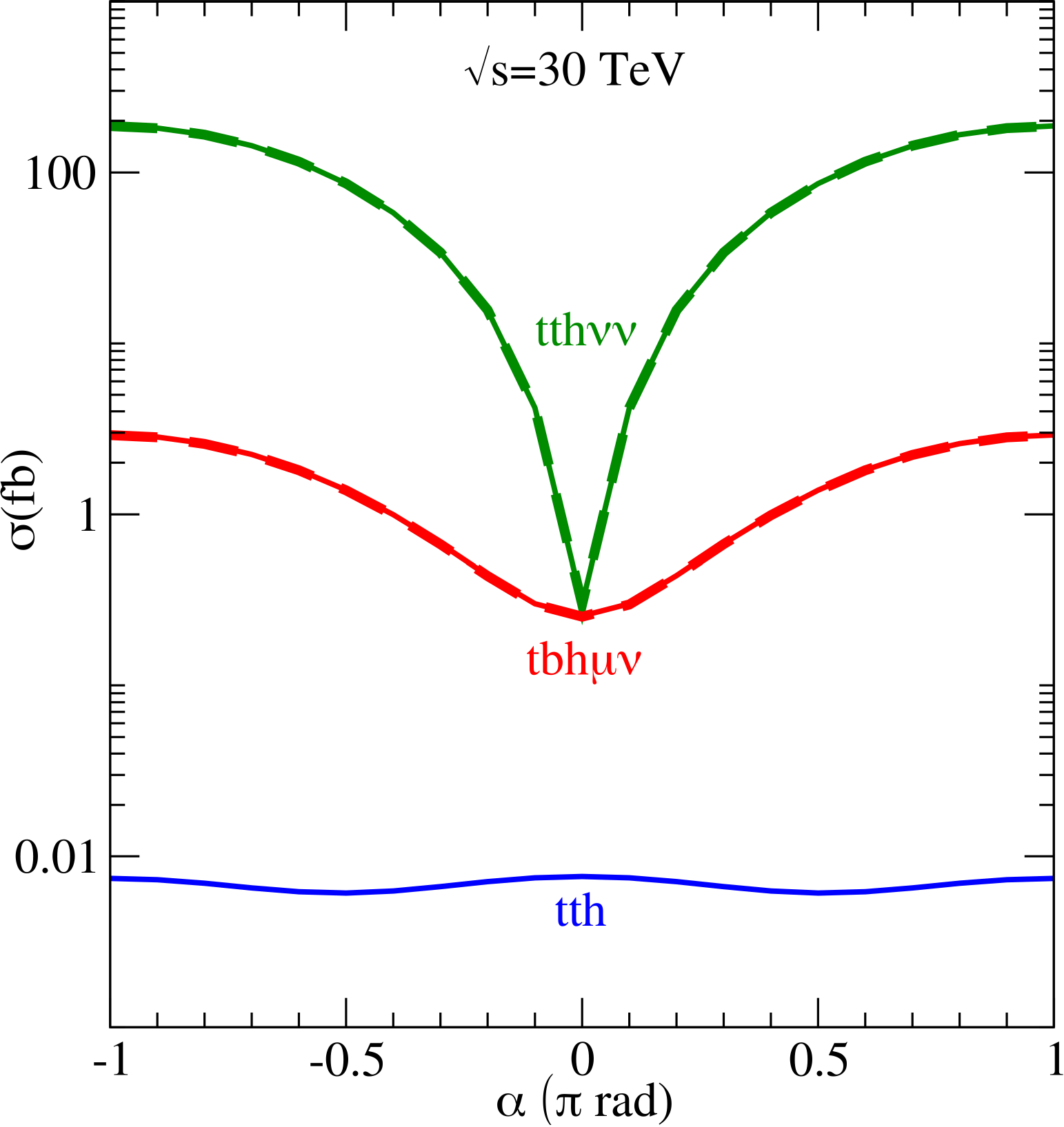}
    \caption{Cross-sections for $\mu^+\mu^-\to t\bar{t}h$ (blue), $t\bar{t}h\nu{\bar\nu}$ (green), and $tbh\mu\nu$ (red) at 1 TeV (left), 10 TeV (middle) and 30 TeV muon colliders (right) with CP violating phase $\alpha$ from $-\pi$ to $\pi$. For $tbh\mu\nu$ a cut of $p_T^\mu>10$ GeV has been applied.  Dashed lines are for the VBF-like contributions.}
    \label{fig:xsCPV}
\end{figure}

In Fig.~\ref{fig:xsCPV}, we show the dependence of the signal cross-sections on the CP-violating angle $\alpha$ for $\sqrt{s}=1$~TeV (left), $\sqrt{s}=10$~TeV (middle), and $\sqrt{s}=30$~TeV (right).  Even though the muon collider is designed for better sensitivity reach for $3$ TeV and above, we start from $1$ TeV to gauge our understanding at lower energy.   It is interesting to note that the dependence on $\alpha$ is significantly different between the signal processes and different collider energies.  Hence, it could be hoped that making a measurement of $\alpha$ at different energies and in different processes could give complementary information.  At $\sqrt{s}=1$ TeV where VBF is subdominant, each signal cross-section shows a similar dependence on $\alpha$.  For $\sqrt{s}=10$ and $30$~TeV, the dependencies of $\alpha$ vary significantly between each process.  While $t\bar{t}h$ has a maximum cross-section at the SM point $\alpha=0$, both $t\bar{t}h\nu\bar{\nu}$ and $tbh\mu\nu$ have minimum cross-sections at $\alpha=0$.  Also, while $t\bar{t}h\nu\bar{\nu}$ and $tbh\mu\nu$ has similar SM cross-sections, as the CP violating phase moves away from zero $t\bar{t}h\nu\bar{\nu}$ becomes the dominant signal.

Now we move onto a collider analysis with a full signal vs. background simulation to extrapolate how well a future muon collider can constrain the CP violating phase.  We consider the semileptonic decay of the top quarks with $h \to b \bar b$.  The backgrounds we include are $b\bar b (g \to t\bar t)$, $t\bar t (g \to b\bar b)$, $t\bar t (Z/\gamma^* \to b\bar b)$, $W^* W^* \to t \bar b \bar t b$, and $t\bar t b \bar b \nu\bar\nu$, where paranthesis are used to indicate gluon, photon, or $Z$ splitting.  We also include detector effects with a Gaussian smearing of jet energies.  Before cuts $t\bar{t}h\bar{\nu}\nu$ and $tbh\mu\nu$ have similar cross-sections.  However, $tbh\mu\nu$ has a very low cut efficiency, so $t\bar{t}h\bar{\nu}\nu$ gives by far the strongest constraints at $\sqrt{s}=10$ and $\sqrt{s}=30$~TeV.

\begin{figure}[!t]
    \centering
    \includegraphics[width=0.55\textwidth]{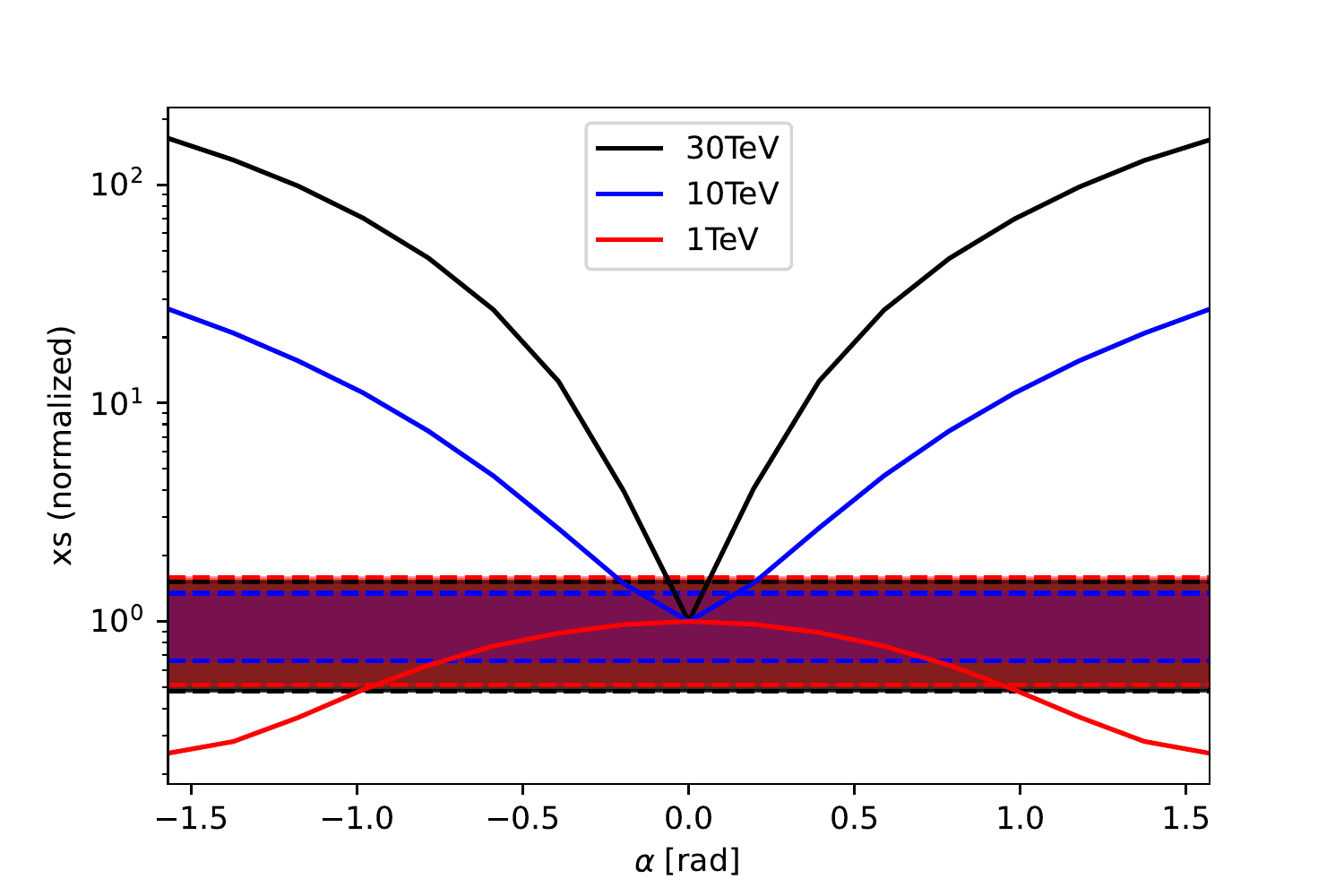}
    \caption{2$\sigma$ exclusion on $\alpha$ at (red) 1 TeV, (blue) 10 TeV and (black) 30 TeV muon colliders with luminosities of 100 fb$^{-1}$, 10 ab$^{-1}$, and 10 ab$^{-1}$, respectively. The solid lines show the combined signal cross-section before cuts normalized to the SM prediction.  The horizontal lines represent the projected bounds on the cross-section normalized to the SM production cross-section for each energy.  These bounds are statistical only. 
    }
    \label{fig:excl}
\end{figure}

In Fig.~\ref{fig:excl}, we show the resulting 95\% CL confidence level bounds that result from the collider analysis, the results of all signal channels are combined.   These bounds are statistical only.  Systematic uncertainties can be added in quadrature.   We overlay results for (red) a 1 TeV muon collider with 100 fb$^{-1}$ of data, (blue) 10 TeV with 10 ab$^{-1}$, and (black) 30 TeV with 10 ab$^{-1}$. The horizontal lines represent the bounds on the cross-section normalized to SM production cross-section for each energy. The bounds on the CP-violating angle are $|\alpha|\lesssim 55^\circ$, $|\alpha|\lesssim 9^\circ$ and $\alpha\lesssim 3^\circ$ for 1, 10 and 30 TeV, respectively.

\section*{Acknowledgements}
R.K.B and D.G. are supported by the United States Department of Energy grant number DE-SC0016013. F.K. is supported by the DFG under Germany’s Excellence Strategy -- EXC 2121 Quantum Universe -- 390833306. K.K., I.M.L., and Y-J.Z. are supported in part by the United States Department of Energy
grant number DE-SC0019474.  M.C. and Y.Z.  are by the State of Kansas EPSCoR grant program. 
C.D. is supported in part by the US DOE under grant No. DE-SC0021447. 

\bibliographystyle{myutphys}
\bibliography{ref}

\end{document}